\documentclass[%
 superscriptaddress,
 preprint,
 showpacs,preprintnumbers,
 amsmath,amssymb,
 aps,
 prb,
]{revtex4-1}

\usepackage{graphicx}
\usepackage{dcolumn}
\usepackage{bm}

\usepackage{upgreek}
\usepackage{amsmath}

\begin{document}


\title{Phase-change metasurfaces for dynamic image display and information encryption}

\author{Tingting Liu}
\email{ttliu@usst.edu.cn}
\affiliation{Institute of Photonic Chips, University of Shanghai for Science and Technology, Shanghai 200093, China}
\affiliation{Centre for Artificial-Intelligence Nanophotonics, School of Optical-Electrical and Computer Engineering, University of Shanghai for Science and Technology, Shanghai 200093, China}

\author{Zhou Han}
\affiliation{Institute for Advanced Study, Nanchang University, Nanchang 330031, China}
\affiliation{Jiangxi Key Laboratory for Microscale Interdisciplinary Study, Nanchang University, Nanchang 330031, China}

\author{Junyi Duan}
\affiliation{Institute for Advanced Study, Nanchang University, Nanchang 330031, China}
\affiliation{Jiangxi Key Laboratory for Microscale Interdisciplinary Study, Nanchang University, Nanchang 330031, China}

\author{Shuyuan Xiao}
\email{syxiao@ncu.edu.cn}
\affiliation{Institute for Advanced Study, Nanchang University, Nanchang 330031, China}
\affiliation{Jiangxi Key Laboratory for Microscale Interdisciplinary Study, Nanchang University, Nanchang 330031, China}

\begin{abstract}
	
Optical metasurfaces enable to engineer the electromagnetic space and control light propagation at an unprecedented level, offering a powerful tool to achieve modulation of light over multiple physical dimensions. Here, we demonstrate a Sb$_{2}$S$_{3}$ phase-change metasurface platform that allows active manipulation of both amplitude and phase. In particular, we implement dynamic nanoprinting and holographic image display through tuning crystallization levels of this phase-change material. The Sb$_{2}$S$_{3}$ nanobricks tailored to function the amplitude, geometric and propagation phase modulation constitute the dynamic meta-atoms in the multiplexed metasurfaces. Using the incident polarizations as decoding keys, the encoded information can be reproduced into a naonprinting grayscale image in the near field and two holographic images in the far field. These images can be switched on and off by taking advantages of the reversible tunability of Sb$_{2}$S$_{3}$ nanostructure between amorphous and crystalline states. The proposed phase-change metasurfaces featuring manifold information and multifold encryption promise ultracompact data storage with high capacity and high security, which suggests an exciting direction for modern cryptography and security applications.

\end{abstract}

\maketitle


\section{\label{sec1}Introduction}

The ability for arbitrary control of optical wavefront is a long sought-after goal in the areas of optical imaging, optical communication, and information processing. Metasurfaces, the optically thin artificially structured materials, provide a powerful platform to manipulate light field over multiple physical dimensions including amplitude, phase, polarization, and orbital angular momentum at subwavelength scale \cite{yu2011light,sun2012gradient,wen2015helicity,wu2018controlling,ren2020complex,gao2021reconstruction}. Through precise design and layout of meta-atoms, the optical response of each meta-atom can be independently and completely controlled, enabling novel phenomena and functionalities for optical device applications such as nanoprinting, holography, flat lens, and vortex beam generators \cite{wang2018broadband,gao2018nonlinear,li2019amplitude,deng2020malus,bao2020minimalist,zheng2021metasurface,zheng2021all}. In the past few years, multifunctional metasurfaces designed to synchronously manipulate two or more properties are attracting enormous research interests due to the great potentials for increasing light integration and information capacity \cite{zhang2017multichannel,jin2018noninterleaved,lim2019holographic,bao2019full,wei2019simultaneous,hu2019trichromatic,wen2020multifunctional,deng2020full, luo2020integrated,liu2021multifunctional,liu2021multifunctional,ren2021non}. One prominent example is the combined nanoprinting holograms in a single metasurface design by simultaneously controlling the amplitude and phase of incident light at a desired wavelength, empowering advanced applications in information display and storage \cite{zhao2019metasurface,zhang2019multichannel,li2020three,zhou2022multifold}. However, most of the multifunctional metasurfaces demonstrated so far are designed to be static with fixed optical responses once they are fabricated, and dynamic multifunctional metasurfaces for the advanced meta-device applications in information encryption have not yet been well addressed. 

Chalcogenide phase-change materials (PCMs) have emerged as an excellent class of active materials to realize dynamic metasurfaces in recent years \cite{zheludev2012metamaterials,wuttig2017phase,ding2019dynamic,delaney2020new,xiao2020active,de2020tunable,mandal2021reconfigurable}. Compared with other counterparts, chalcogenide PCMs whose optical refractive index can be remarkably modified between amorphous and crystalline states, show unique advantages such as ultrafast reversible switching (10 $\sim$ 100 ns) between the two stable states, reconfigurable phase change potentially up to 10$^{15}$ cycles and adaptability with CMOS fabrication technology. With these unparalleled properties, chalcogenide PCMs have shown their potential in metasurfaces for optical switching, beam steering, photonic spin-orbit interactions, and optical computing \cite{de2018nonvolatile,feldmann2019all,zheng2020nonvolatile,zhang2020multistate,delaney2021nonvolatile}. In early works, PCMs usually appear as a spacer layer combined into the plasmonic or dielectric nanostructures for the sake of easy fabrication, and the dynamic modulation originates from tuning local environments \cite{gholipour2013all,lee2017holographic,zhou2020switchable,zhou2020imaging,zhou2020optical,zhu2020realization,mao2020reversible,abdelraouf2021multistate,liu2021tuning}.  Most recently, the pioneered experimental works developed the nanofabrication techniques to pattern dielectric nanostructures directly out of PCMs, offering the possibility to tune the meta-atoms made of PCMs itself for nonvolatile and energy-efficient active devices. For example, the structured GST metasurfaces are reported for bifocal zoom lensing, vortex switching, holographic image switching, cryptography, and so on across the infrared region\cite{wang2016optically,chu2016active,tian2019active,li2019active,choi2019metasurface,choi2021hybrid,shalaginov2021reconfigurable}. And in the visible window, a series of novel PCMs materials including Sb$_{2}$S$_{3}$, Sb$_{2}$Se$_{3}$, and GeSe$_{3}$ have been recently successfully employed as nanostructures to realize the dynamically tunable resonances for high-resolution color and beam steering\cite{dong2019wide,lu2021reversible,chen2021multifunctional,hemmatyar2021advanced,moitra2022tunable}. However, the more challenging modulation of optical wavefront over multi-dimensional physical using chalcogenide PCM metasurfaces remains to be demonstrated. 

In this paper, we demonstrate a complete and tunable wavefront control in the visible using a multifunctional metasurface composed of chalcogenide PCM meta-atoms. As an example, we implement integration of nanoprinting and holographic image display for multifold information multiplexing, and realize the multi-level modulation of these images using the refractive index tunability of nanostructured Sb$_{2}$S$_{3}$. Compared with previous phase-change metasurfaces controlling only one property of light, the proposed Sb$_{2}$S$_{3}$ metasurfaces are designed to actively control phase and amplitude simultaneously for multiple independent channels. In the design, the three kinds of images are independently stored in a metasurface composed of one type of Sb$_{2}$S$_{3}$ nanobrick. Using the incident polarizations as decoding keys, the encoded information can be reproduced into a naonprinting grayscale image in the near field and two holographic images in the far field. In particular, by tuning the crystallization level of Sb$_{2}$S$_{3}$, these functions can be controllable in both near and far fields, affording more degrees of freedom for visible information storage and encryption with advantages of crosstalk-free and ultra-compactness.
 
\section{\label{sec2}Working principle}

\begin{figure*}[htbp]
\centering
\includegraphics
[scale=0.8]{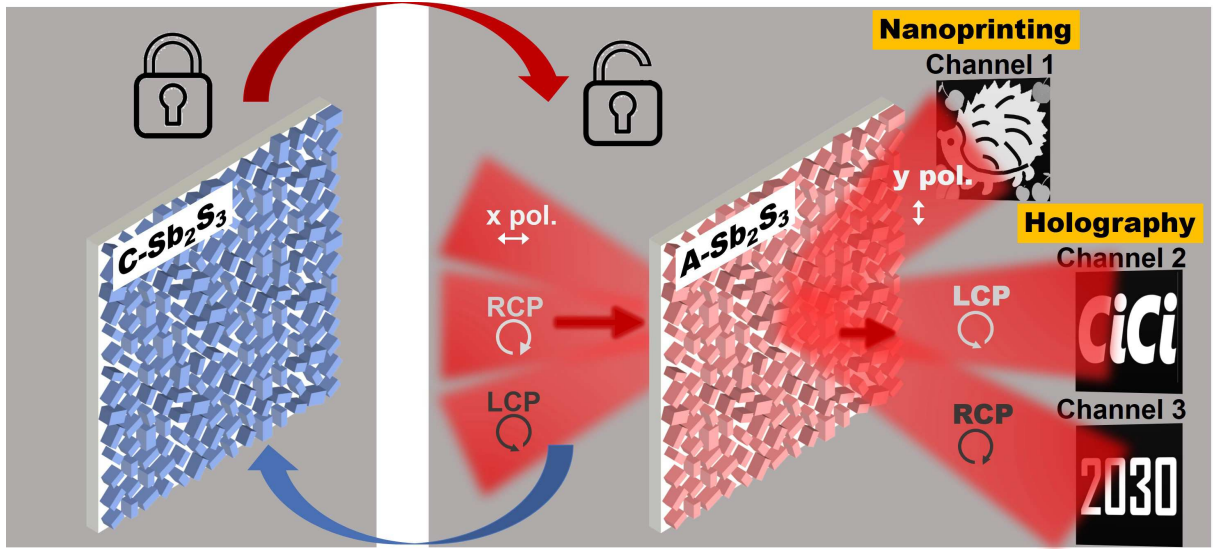}
\caption{\label{fig1} Schematic of dynamic multifunctional metasurfaces based on nanostructured phase-change material. Each  Sb$_{2}$S$_{3}$ nanobrick with optimized size and orientation plays the role of a pixel of diffractive elements that generates the required local amplitude and phase profiles under the normally incident light. The three independent images can be displayed under different polarization combinations of light and dynamic image display can be realized using tunable phase states of  Sb$_{2}$S$_{3}$. When the $x$-polarized light is incident on the designed metasurface, a grayscale nanoprinting image can be decoded right at the metasurface plane along the y-polarized transmission light (Channel 1). Under the circularly polarized (CP) incident light including RCP and LCP, the opposite handedness CP light, i.e. LCP and RCP are collected for decoding the two far-field holographic images (Channel 2 and 3). By reversibly tuning between the amorphous and crystalline states of Sb$_{2}$S$_{3}$, image information can be displayed or hidden for dynamic display and encryption.}
\end{figure*}

The concept of structured Sb$_{2}$S$_{3}$ metasurfaces for dynamic image display and information encryption is illustrated in Fig. 1. The proposed metasurfaces can work as a nanoprinting meta-image display device and a holography device at the same time. Under the incident linearly polarized (LP) light, the metasurfaces can produce a continuous grayscale nanoprinting image in the near field when inserted into an orthogonal-polarization optical setup. Under the left-handed circularly polarized (LCP) and right-handed circularly polarized (RCP) laser light, the two different high-quality holographic images are projected in the far field, respectively. The image information of all the three channels are recorded using the Sb$_{2}$S$_{3}$ meta-atoms with varied dimensions and orientations, and the three information channels are independent to carry different images. These display functions can be readily switched repeatedly and quickly between the amorphous (A-Sb$_{2}$S$_{3}$) and crystalline states (C-Sb$_{2}$S$_{3}$) of Sb$_{2}$S$_{3}$. In the amorphous state, the metasurfaces are characterized by amplitude and phase profiles that encode these images information of three channels. During the phase transition of  Sb$_{2}$S$_{3}$ nanostructures to the full crystallinity, both amplitudes and phases gradually vary and finally no valuable information can be read. 

The proposed dynamic multifunctional metasurfaces encode three channels of completely independent information into a single piece of design, which is essentially different from the previously reported functional metadevices \cite{li2018addressable,li2019reconfigurable,jia2020composite,hu2021electrically,dong2022terahertz}. Each Sb$_{2}$S$_{3}$ meta-atom serves as the minimum operating unit and contributes to the wavefront control. Especially, unlike the supercell and multilayer design strategies for multi-manipulations of wavefront, only one type of meta-atom is adopted to realize these independent functions. The single-cell metasurface with subwavelength pixel size is realized for actively engineering both amplitude and phase of light. Hence the proposed metasurface possesses ultrahigh information density and avoids the fundamental limitations such as crosstalk and pixel loss. Also, the tunability of Sb$_{2}$S$_{3}$ meta-atoms makes it possible to realize multistate, reversible, nonvolatile, and ultrafast manipulation of the image display from the meatsurfaces. More importantly, the dynamic multifunctionality lead to more freedoms and higher security for information encryption. For the example shown in Fig. 1, the identity information of a hedgehog is encoded in the three independent channels. His photograph appears in the near field under LP light, while his name ‘CiCi’ and birth year ‘2030’ are encrypted in the far-field holographic image projections under the CP light. Furthermore, with the phase change of Sb$_{2}$S$_{3}$ to the crystalline state, CiCi’s identity can be completely hidden for a full encryption. As a result, the proposed metasurfaces show the advantages of multi-functionality, high information capacity, and security, offering a prototype  for  potential applications such as ultracompact image displays, high-density information storage, information encryption, etc.

To achieve the dynamic multifunctionality above, we combine the amplitude modulation in the near field governed by Malus’ law with the phase modulation in the far field based on the geometric phase and propagation phase into a single structured Sb$_{2}$S$_{3}$ metasurface. Without the loss of generality, the transmissive anisotropic Sb$_{2}$S$_{3}$ nanobricks are adopted to encode the three independent information channels. On one hand, the anisotropic nanobricks can serve as a nano half-waveplate that enable to vary the polarization direction of LP light, and hence the amplitude manipulation of the metasurface can be realized by adjusting the orientations of the nanobricks. On the other hand, such nanobricks can introduce distinct phase shifts for the two orthogonal polarizations along its fast and slow axis, and the modulation of geometric phase can also be realized by their orientations. Through the precise design using the connection, the wavefront modulation over the amplitude and phase space can be possible.

In the following, the optical response of a single Sb$_{2}$S$_{3}$ meta-atom are firstly considered to present the operational mechanism of the proposed metasurface. Each Sb$_{2}$S$_{3}$ nanobrick can be regarded as a Jones matrix connecting the input field to the output field, as $E_{o}=T \cdot E_{i}$. By rotating the meta-atoms by angle $\theta$ with respect to the $x$ direction, the transmission Jones matrix can be decomposed into the multiplication of the rotation matrix $R(\theta)$ and a diagonal matrix in the Cartesian basis as
\begin{equation} \label{eq1}
T=R(\theta) 
\begin{bmatrix} 
t_l & 0  \\
0   & t_s
\end{bmatrix} 
R(-\theta) 
=\begin{bmatrix} 
\cos\theta & -\sin\theta  \\
\sin\theta  & \cos\theta
\end{bmatrix} 
\begin{bmatrix} 
t_l & 0  \\
0   & t_s
\end{bmatrix}
\begin{bmatrix} 
\cos\theta & \sin\theta  \\
-\sin\theta  & \cos\theta
\end{bmatrix},
\end{equation}
where $t_l$ and $t_s$ are the complex transmission coefficients along the long and short axes of the nanobrick, respectively. Thus the Jones matrix can be expressed by 
\begin{equation}\label{eq2}
T=\begin{bmatrix} 
t_l\cos^{2}\theta+t_s\sin^{2}\theta & \sin\theta\cos\theta(t_l-t_s)  \\
\sin\theta\cos\theta(t_l-t_s)   & t_l\cos^{2}\theta+t_s\sin^{2}\theta
\end{bmatrix}.
\end{equation}

\subsection{\label{sec2.1}Amplitude modulation in the near field}
When the nanobricks are inserted between a bulk-optic polarizer and an analyzer that control the light’s polarization directions, the Jones vector of the transmission light can be obtained
\begin{equation}\label{eq3}
E_{o}=\begin{bmatrix} 
\cos^{2} \alpha_{2} & \sin\alpha_{2} \cos\alpha_{2}  \\
\sin\alpha_{2} \cos\alpha_{2}   & \sin^{2} \alpha_{2}
\end{bmatrix}
T
\begin{bmatrix} 
\cos \alpha_{1}   \\
\sin\alpha_{1}  
\end{bmatrix},
\end{equation}
where $\alpha_{1}$, $\alpha_{2}$ are the angles between the transmission axis directions of the bulk-optic polarizer and analyzer, respectively. Therefore, a general expression of the output transmission light is written as
\begin{equation}\label{eq4}
E_{o}=(\frac{t_l+t_s}{2} \cos(\alpha_{2}-\alpha_{1})+\frac{t_l-t_s}{2}\cos(2\theta-\alpha_{2}-\alpha_{1}))
\begin{bmatrix} 
\cos \alpha_{1}   \\
\sin\alpha_{1}  
\end{bmatrix}.
\end{equation}
For a simple case of incident light polarized along the $x$ axis and the analyzer for transmission light along the $y$ axis, the angles can be set as $\alpha_{1}=0$ and $\alpha_{2}=\pi/2$, respectively. And the output light intensity is calculated as
\begin{equation}\label{eq5}
I=(\frac{t_l-t_s}{2})^{2} \sin^{2}(2\theta).
\end{equation}
The first term is usually considered as the cross-polarization conversion efficiency of Sb$_{2}$S$_{3}$ nanobricks, i.e. $T_{cross}= |(t_l-t_s)/2|^2$, which only depends on the phase difference between the fast and slow axis directions of the nanostructure. As a result, the near-field amplitude can be continuously manipulated by only adjusting the rotation angle $\theta$ of Sb$_{2}$S$_{3}$ nanobricks. It is also noted that the concept of orientation degeneracy, i.e., the one-to-many mapping relationship between the light amplitude and rotation angle can also be well explained here. As shown in Eq. (\ref{eq5}), the four orientations including $\theta$, $\pi/2-\theta$, $\pi/2+\theta$, and $\pi-\theta$ can generate equal amplitude. Such a one-to-four mapping offers new degrees of freedom in designing Sb$_{2}$S$_{3}$ nanobricks. 

\subsection{\label{sec2.2}Phase modulation in the far field}
When a LCP or a RCP light is employed to normally illuminate Sb$_{2}$S$_{3}$ nanobricks, i.e. $E_{i}=\frac{1}{\sqrt{2}} [1 \quad \delta i]^{T}$  where $\delta=1$ or $-1$ corresponds to the LCP or RCP light, the output transmission light can be calculated as
\begin{equation}\label{eq6}
E_{o}=T \cdot 
\frac{1}{\sqrt{2}}
\begin{bmatrix} 
1   \\
\delta i  
\end{bmatrix}
=\frac{t_l+t_s}{2}
\begin{bmatrix} 
1   \\
\delta i  
\end{bmatrix}
+\frac{t_l-t_s}{2}
e^{i2\theta}
\begin{bmatrix} 
1   \\
-\delta i  
\end{bmatrix}.
\end{equation}
For the output field, the first term is regarded as the co-polarized transmission component, and the second term carries the cross-polarized component with the opposite handedness as the incident light. The polarization conversion efficiencies for the two components can be defined by $T_{co}=|(t_l+t_s)/2|^{2}$ and $T_{cross}= |(t_l-t_s)/2)|^2$. For the cross-polarized component, the phase of the transmission light $\varphi$ includes the spin-dependent geometric phase and the spin-independent propagation phase 
\begin{equation}\label{eq7}
\varphi=\phi\pm\Psi.
\end{equation}
Here $\phi$ represents the propagation phase of complex $(t_l-t_s)/2$, and $\Psi=2\theta$ represents the part of geometric phase. Thus, the full phase modulation covering 0 to $2\pi$ can be achieved by the reasonable combination of both propagation and geometric phases, which originates from the design of meta-atoms with various dimensions and orientations. Under the CP light with different helicities, the combined phases $\varphi$ can be independently modulated by the transmissive metasurface, which provides the two distinct channels for the far-field holograph projections. 

\subsection{\label{sec2.3}Dynamic modulation in both the near and far fields}
Employing patterned PCMs in metasurface design is an excellent strategy to obtain a dynamic wavefront control. Despite the great success of GST in reconfigurable meta-devices within the infrared regime, the unconventional Sb$_{2}$S$_{3}$ material with a large bandgap of 1.70$\sim$2.05 eV shows a relatively low absorption in the visible spectrum around 600 nm \cite{lu2021reversible,lu2022reconfigurable}. The refractive index of the Sb$_{2}$S$_{3}$ remains high around $\sim$3.5 in the visible, and it is comparable to Si, TiO$_{2}$ in recent reports, making the material a good candidate for all-dielectric metasurfaces. In this sense, the wavefront modulation of the metasurfaces can be achieved by properly designing the single type of Sb$_{2}$S$_{3}$ nanobrick to simultaneously satisfy the aforementioned phase and amplitude conditions. Using appropriate thermal, electric or optical stimulus, the reversible multilevel switching of Sb$_{2}$S$_{3}$ nanostructures can be achieved between the amorphous and crystalline states. During the phase transition to crystalline, the refractive index increases with $\Delta n \sim 0.5$ and the large contrasts can simply vary the transmission response. Meanwhile, the light extinction coefficient $k$ also becomes apparent to increase absorption loss. These would disorder or erase the designed functions of metasurface, leading to the dynamic switching of the multifunctionality.

\section{\label{sec3}Results and Discussion}

\begin{figure*}[htbp]
	\centering
	\includegraphics
	[scale=0.6]{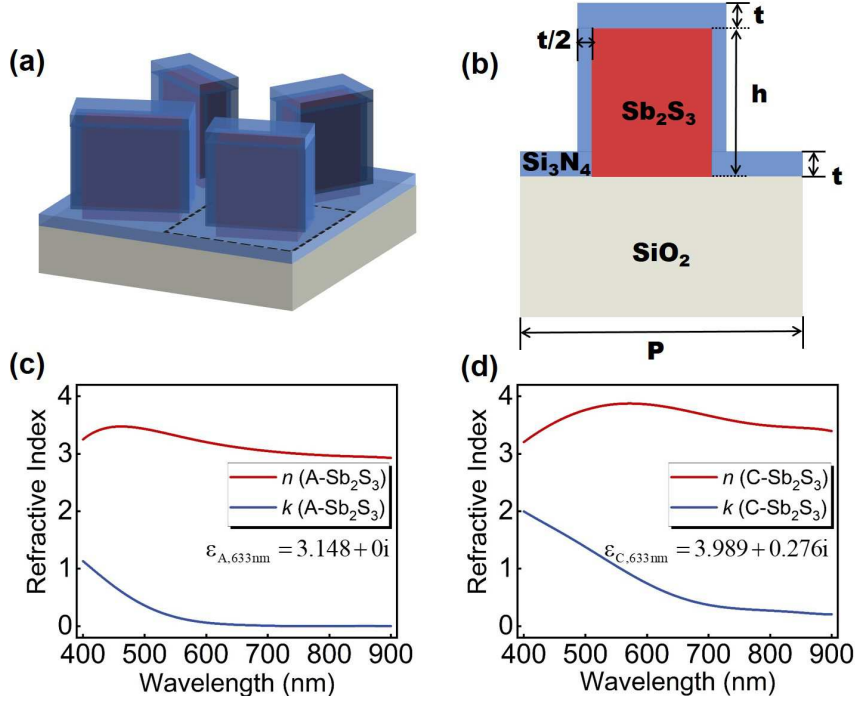}
	\caption{\label{fig2}  Illustration of the phase-change metasurfaces composed of Sb$_{2}$S$_{3}$ nanobricks and the material optical properties.  (a) Design of the Sb$_{2}$S$_{3}$ nanobrick structure on a silica substrate encapsulated by Si$_{3}$N$_{4}$ film. (b) The front view of a unit cell in the proposed metasurface. The refractive index $n$ and the extinction coefficient $k$ of Sb$_{2}$S$_{3}$ for both (c) amorphous and (d) crystalline states. }
\end{figure*}

For a proof-of-concept demonstration, the phase-change metasurfaces sketched in Fig. 2(a) and (b) are designed to the implement the dynamic nanoprinting and holographic image display. Sb$_{2}$S$_{3}$ nanobricks are the building blocks to construct the desired metasurface on the silica substrate. The periods in both $x$ and $y$ directions are $P= 400$ nm, and the height of nanobricks are $h=500$ nm. The most suitable dimensions and orientations of meta-atom can be selected for the three information channels with extra design freedoms. For the sake of practical fabrication, non-reactive dielectric materials like Si$_{3}$N$_{4}$ or the conventional resists such as PMMA are usually employed to protect the soft material Sb$_{2}$S$_{3}$ \cite{lu2021reversible,moitra2022tunable}. Here a Si$_{3}$N$_{4}$ film with thickness of $t=70$ nm is used to encapsulate the Sb$_{2}$S$_{3}$ nanobricks in the design. The Si$_{3}$N$_{4}$ encapsulation not only prevents the heating damage of Sb$_{2}$S$_{3}$ materials during the reversible tuning of crystallization and amorphization, such as the sulfur loss through evaporation, but also effectively mimics a free-space environment for stable performance of nanostructures. Fig. 2(c) and (d) present the optical constants of Sb$_{2}$S$_{3}$ for the two states \cite{lu2021reversible}. The refractive index $n$ is well above 3 in the entire visible regime, and the C-Sb$_{2}$S$_{3}$ has the larger $n$ and the extinction coefficient $k$.  

The metasurfaces are targeted to implement tunable multiple tasks that merge the three kinds of independent images including one nanoprinting and two holographic images into a single Sb$_{2}$S$_{3}$ nanostructure. Fig. 3(a) presents the working mechanism of the three channels. Based on the amplitude and phase modulation strategies, the unit cell of metasurface consisting of only one Sb$_{2}$S$_{3}$ nanobrick is designed with optimized length ($L$), width ($W$), and orientation angle ($\theta$). Herein we choose the working wavelength of 633 nm and carry out numerical calculations with finite-difference-time-domain (FDTD) method via commercial package FDTD Solutions. The transmission amplitude in the near field is firstly considered. When the $x$-polarized light is normally incident from the substrate side of the metasurface, the simulated amplitude of the $y$-polarized transmission amplitude is shown in Fig. 3(c). It can be clearly observed that the relation between the amplitude and the orientation angle is in accord with the line of $\sin^{2}(2\theta)$, as illustrated in Eq. (\ref{eq5}). The stimulated geometric phase $\Psi$ versus orientation angle $\theta$ are also presented in Fig. 3(d), where $\Psi$ is calculated using the stimulated cross circularly polarized transmission response. The geometric phase obeys the relation of $\Psi=2\theta$, implying that it is only determined by the orientation angle $\theta$ and incident helicity of light. Due to the extra freedom of orientation degeneracy, the angle of each nanobrick can be chosen to be closest to the required geometric phase from the one-to-four mapping relation. 

\begin{figure*}[htbp]
	\centering
	\includegraphics
	[scale=0.80]{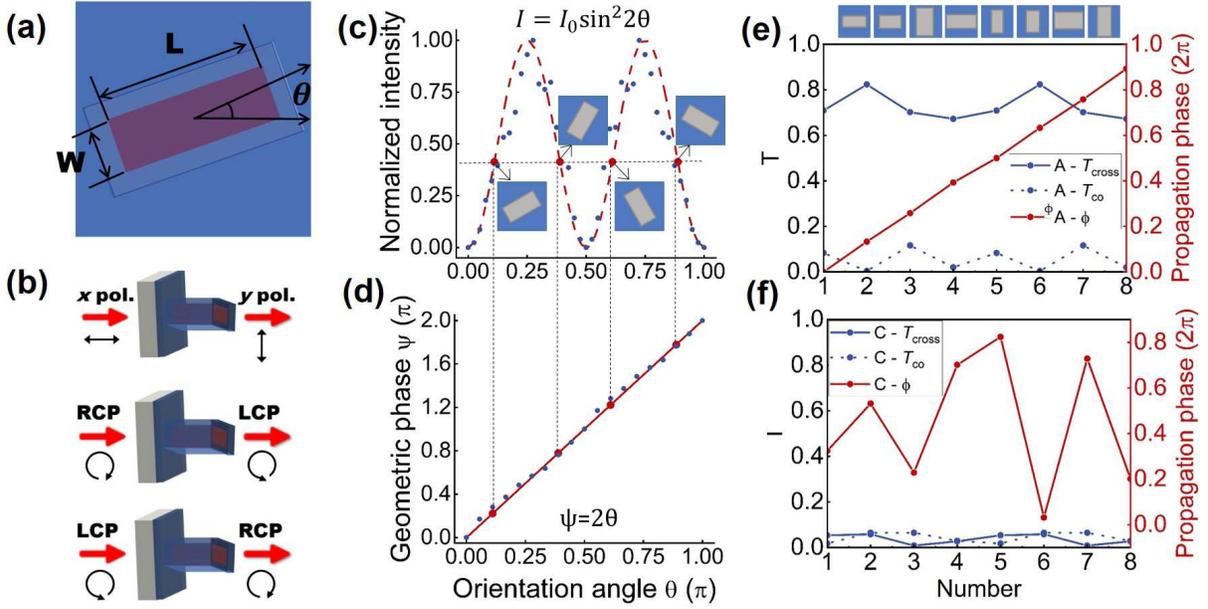}
	\caption{\label{fig3} Amplitude and phase modulation for three-channel Sb$_{2}$S$_{3}$ metasurfaces. (a) The unit cell consisting of only one Sb$_{2}$S$_{3}$ nanobrick is designed with optimized length ($L$), width ($W$), and orientation angle ($\theta$) for the independent nanoprinting and holographic image display dependent on incident polarizations. (b) The working mechanism of the three channels. (c) The normalized amplitude of the $y$-polarized transmission amplitude as a function of the orientation angles $\theta$. (d) The geometric phase $\Psi$ as a function of orientation angle $\theta$. The cross-polarized amplitude efficiency T$_{cross}$ and propagation phase $\phi$ distributions of the selected eight meta-atoms with various lengths and widths of nanobricks (e) A-Sb$_{2}$S$_{3}$ and (f) C-Sb$_{2}$S$_{3}$.}
\end{figure*}

Then the propagation phase is introduced to establish the holographic information channels together with geometric phase as Eq.(\ref{eq7}) shows. Owing to the sensitivity to the size of meta-atoms, the cross polarization component under CP incidence are calculated in simulations for different combinations of length $L$ and width $W$ of Sb$_{2}$S$_{3}$ nanobricks. At the wavelength of interest, the cross-polarized amplitude $T_{cross}$ and propagation phase $\phi$ of the component are studied for both A-Sb$_{2}$S$_{3}$ and C-Sb$_{2}$S$_{3}$. In both states, $\phi$ of the component cover 0 $\sim 2\pi$. After the full crystallinity, larger extinction ration $k$ of C-Sb$_{2}$S$_{3}$ leads to increasing absorption loss and thus much lower performance of $T_{cross}$. These characteristics guide the optimization design and help us to find desired meta-atoms. Specifically, we carefully design the sizes of eight Sb$_{2}$S$_{3}$ nanobricks with an incremental propagation phase of $\approx\pi/4$ and high amplitudes of $T_{cross}$ in the amorphous state, but low transmission performance in the crystalline state. The simulated results of the selected eight meta-atoms at LCP incidence are presented for the two states in Fig. 3(e) and (f), respectively. In contrast to the step phase and uniform transmission for A-Sb$_{2}$S$_{3}$, the selected meta-atoms show the random phase distributions and extremely low efficiency for C-Sb$_{2}$S$_{3}$. When we combine geometric phase with propagation phase to generate the phase delays for the incident LCP and RCP light, the two independent information channels are established for phase-only holographic images. Due to the extremely different phase and amplitude distributions in the eight-step propagation phases, the coded image information of the two channels in A-Sb$_{2}$S$_{3}$ metasurface cannot be decoded in the C-Sb$_{2}$S$_{3}$ case. 

\begin{figure*}[htbp]
	\centering
	\includegraphics
	[scale=0.80]{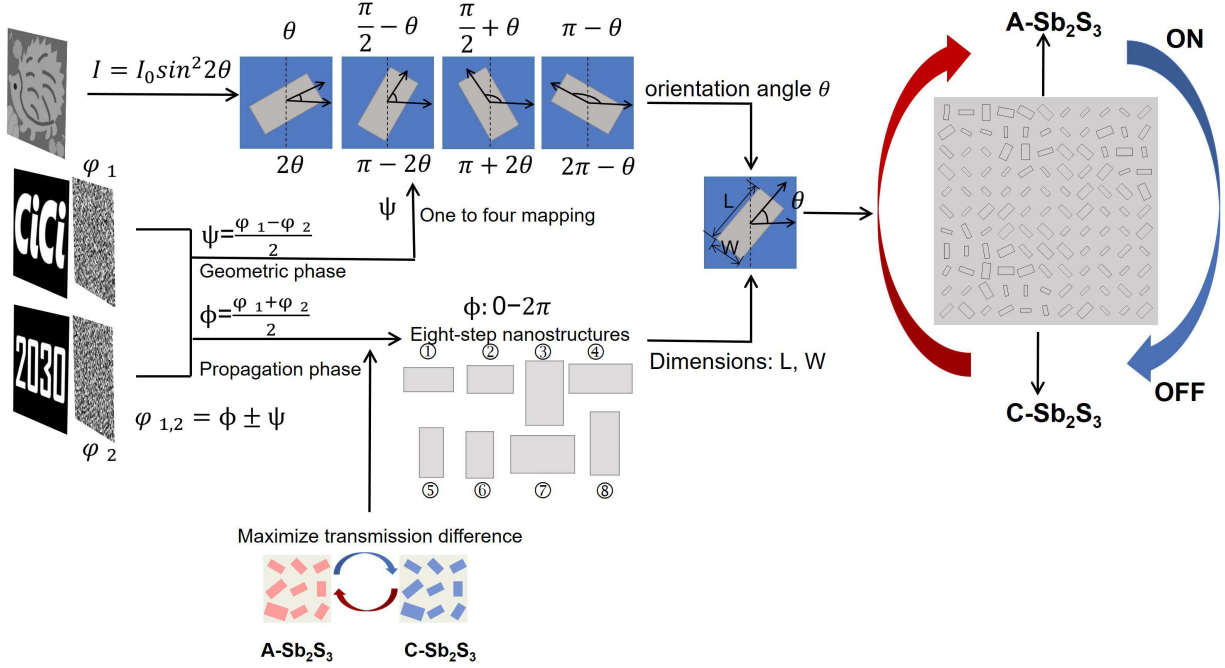}
	\caption{\label{fig4} Flowchart of designing the dynamic multifunctional Sb$_{2}$S$_{3}$ metasurfaces for switchable nanoprinting and holography. A grayscale hedgehog’s photograph is selected as the target nanoprinting image, and two different images containing the name ‘CiCi’ and birth year ‘2030’ are selected as the target holographic images. Firstly, the transmission amplitude distribution of target grayscale image is used to obtain the four selectable orientation angles $\theta$. Secondly, the phase distributions $\varphi_{1}$ and $\varphi_{2}$ of the two target holographic images are obtained by GS algorithm. And the geometric and propagation phases are calculated by $\phi=(\varphi_{1}+\varphi_{1})/2$  and $\Psi=(\varphi_{1}-\varphi_{1})/2$, respectively. Thridly, $\theta$ of the nanobrick at this pixel is selected from the four candiadates which is closet to half of geometric pahse, and its length and width are determined by a lookup approach from the preset eight meta-atom database. At last, the meta-atoms are arranged pixel-by-pixel into a single metasurface. }
\end{figure*}

For demonstration, the identity information of the hedgehog are employed as the target images for the three channels, including his grayscale photograph as the near-field nanoprinting image, his name ‘CiCi’ and birth year ‘2030’ as the two far-field holographic images. The design flowchart is illustrated in Fig. 4. The design of the three-channel metasurface is essentially an inverse problem to find the relationship between ($T_{cross}$, $\phi$, $\Psi$) from the three distinct images and ($\theta$, $L$, $W$) of the Sb$_{2}$S$_{3}$ nanobricks. At first, the grayscale image of hedgehog’s photograph is encoded into the orientations of nanobricks by $I=\sin^2 (2\theta)$. Owing to the orientation degeneracy, four angles are the candidates for the desired amplitude, including $\theta$, $\pi/2-\theta$, $\pi/2+\theta$, and $\pi-\theta$. Secondly, the phase profiles for the two hologram images, $\varphi_{1}$ and $\varphi_{2}$ are calculated by Gerchberg-Saxton (GS) algorithm in a computer-generated hologram (CGH). According to the phase relationships in Eq. (\ref{eq7}), the propagation and geometric phase can be expressed by $\phi=(\varphi_{1}+\varphi_{1})/2$  and $\Psi=(\varphi_{1}-\varphi_{1})/2$, respectively. Using the relationship between geometric phase $\Psi$ and the rotation angles $\theta$, the most suitable angle could be selected from the four candidates. As we stated above, the propagation phase can be encoded by nanobrick structural parameters including the $L$ and $W$. Through the lookup approach from the eight meta-atom database, the nanobrick size at each pixel can be easily determined by the propagation phase. As a result, the Sb$_{2}$S$_{3}$ nanobricks with various sizes and orientations are arranged by pixel into an array, encoding the three target images into a single metasurface platform. Moreover, since the $L$ and $W$ of the nanobricks for encoding eight-step propagation phase is determined through maximizing the differences of optical responses between A-Sb$_{2}$S$_{3}$ and C-Sb$_{2}$S$_{3}$, the designed Sb$_{2}$S$_{3}$ metasurface is expected to possess the dynamic switching functionality.

\begin{figure*}[htbp]
	\centering
	\includegraphics
	[scale=0.80]{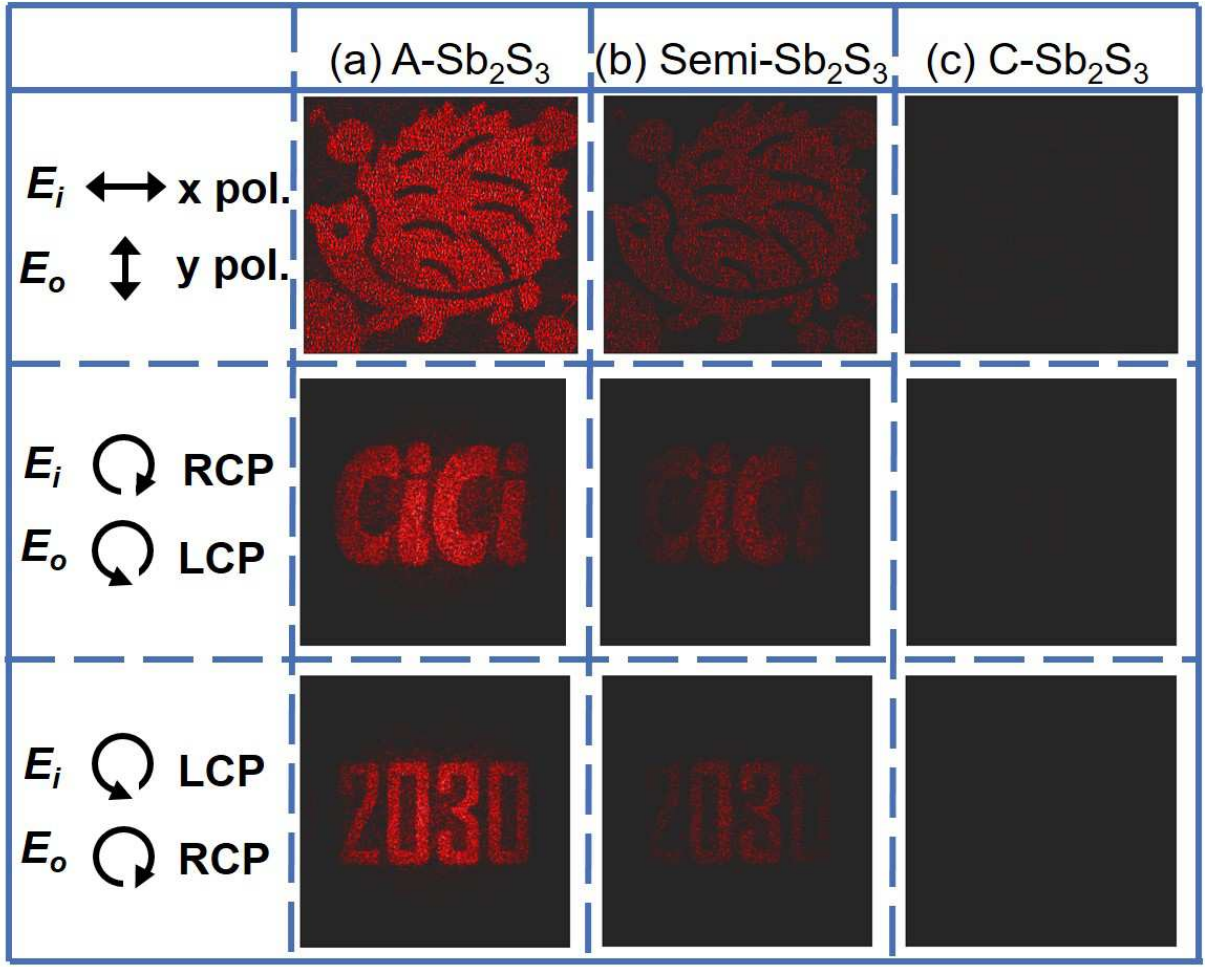}
	\caption{\label{fig5} Demonstration of the multi-level tunable image display for information encryption. Under corresponding polarization combinations, nanoprinting and holographic images of the designed phase-change metasurfaces in (a) amorphous state, (b) semi-crystalline state, and (c) crystalline state of Sb$_{2}$S$_{3}$. }
\end{figure*}

By arranging the designed Sb$_{2}$S$_{3}$ nanobricks, the phase-change metasurfaces can be realized for dynamic nanoprinting and holographic image display. The metasurfaces consisting of $150\times150$ pixel numbers with overall size of $60\times60$ um is employed for the full-wave FDTD simulations. As shown in Fig. 5, the simulation results agree with the preset dynamic multifunctional design. In the amorphous state, when the $x$-polarized light is incident, the grayscale nanoprinting image in the near field can be seen under the $y$ polarization channel. Meanwhile, the two different Fourier holographic images in the far field can be observed in the LCP (RCP) channel converted from the incident RCP (LCP), respectively. In this sense, the encoded images in the three channels are all polarization related and they can only be decoded under the specific polarization combinations.

As the phase transition to full crystallinity, the effective dielectric constant of chalcogenide PCM depends on the crystalline fraction $m$ according to the Lorenz-Lorentz relation \cite{tian2019active,liu2021tuning,meng2021high}, 
\begin{equation}
\frac{\varepsilon_{eff}-1}{\varepsilon_{eff}+2}
=m\times\frac{\varepsilon_{C-Sb_{2}S_{3}}-1}{\varepsilon_{C-Sb_{2}S_{3}}+2}
+(1-m)\times\frac{\varepsilon_{A-Sb_{2}S_{3}}-1}{\varepsilon_{A-Sb_{2}S_{3}}+2},
\end{equation}
where $\varepsilon_{A-Sb_{2}S_{3}}$ and $\varepsilon_{C-Sb_{2}S_{3}}$ denote the wavelength-dependent permittivity of A-Sb$_{2}$S$_{3}$ and C-Sb$_{2}$S$_{3}$, respectively. The increasing crystalline fraction $m$ from 0 to 1 implies the intermediate hybridization state transformed from amorphous to crystalline states of Sb$_{2}$S$_{3}$, giving rise to the multi-level changes of optical constants. Thus, the multi-level wavefront control in the phase-change metasurface can be expected. Herein we select the intermediate state with $m$=0.5 for the demonstration of multi-level control, and it is referred to as semi-crystalline state of Sb$_{2}$S$_{3}$ (semi-Sb$_{2}$S$_{3}$). Then the permittivity of semi-Sb$_{2}$S$_{3}$ is calculated as 3.51+i 0.10 for the concerned wavelength of 633 nm. In such state, the two holographic images under CP light, and the nanoprinting image in an orthogonal-polarization optical path are displayed in Fig. 5(b). In comparison with A-Sb$_{2}$S$_{3}$, the semi-Sb$_{2}$S$_{3}$ shows slightly large $n$ and $k$. Hence, the designed images are preserved and the transmission intensity is reduced, as observed by the results in Fig. 5(b). After the full crystallinity of C-Sb$_{2}$S$_{3}$, $n$ is at a relatively lager value than A-Sb$_{2}$S$_{3}$ with $\Delta n \sim$0.8 and $k$ increases to $\sim$0.3. The propagation phases for holographic channels are changed, disordering the preset phase profile for A-Sb$_{2}$S$_{3}$ metasurface. More importantly, the increased absorption simply erases the images of three channels in the visible regime, reducing the transmission to $\leq$1/10. As shown in Fig. 5(c), the holographic and nanoprinting images are hidden for C-Sb$_{2}$S$_{3}$. Interestingly, the transition between amorphous and crystalline states is reversible. With these unique properties, the phase-change metasurfaces can be dynamically and reversibly switched between on and off states by adjusting the material’s crystalline fraction under external stimulus.  

Based on the results on the dynamic nanoprinting and holographic image displays, it is confirmed that the phase-change metasurfaces offer an excellent platform to achieve the complete wavefront modulation across amplitude and phase spaces and dynamic multifunctionalities. In terms of information transmission, such a design provides the abundant degrees of freedom for a wide variety of applications associated with optical information encryption. In particular, there are multifold encryption mechanisms in the proposed single metasurface. Since the three independent images in the single metasurface can only be observed under the corresponding polarization states, the decoded polarization combinations can be considered as the security keys, making these images as encrypted information carries. The other mechanism lies in employing the reversible tunability of the phase change materials to reveal and hide information, for example, the image information coded by nanostructured Sb$_{2}$S$_{3}$ can only be captured at amorphous states but not observed at crystalline states. Therefore, the proposed phase-change metasurface provides an advanced encryption strategy to significantly increase the information security. 

\section{\label{sec4}Conclusions}

In conclusion, we present a methodology to realize the dynamic multifunctional metasurfaces by employing nanostructured chalcogenide PCM as building blocks. For demonstration, the tunable and simultaneous amplitude and phase modulations are realized in the phase-change metasurfaces composed of Sb$_{2}$S$_{3}$ nanobricks in the visible regime. By merging the amplitude, geometric and propagation phase modulation into a single-celled Sb$_{2}$S$_{3}$ metasurface, a near-field nanoprinting image and two far-field holographic images are displayed and actively switched. The main idea of our design lies in recording the different images of three channels into one phase-change metasurface by taking advantages of the design freedom among multiple channels and especially the reversible tunability of Sb$_{2}$S$_{3}$ nanostructures. The numerical results validate that three channels of information can be coded in the proposed metasurface with high information capacity and security. Decoding these image information requires the certain polarization combinations. Importantly, owing to the tunability of Sb$_{2}$S$_{3}$ nanostructures, the multiplexing information channel can be switched. With the advantage of multifunctionality, compactness, and high-security, the proposed phase-change metasurfaces exhibit considerable potential in active meta-devices, and will find wide applications in information storage and encryption.

\begin{acknowledgments}	
	
This work is supported by the National Natural Science Foundation of China (Grants No. 11947065 and No. 61901164), the Natural Science Foundation of Jiangxi Province (Grant No. 20202BAB211007), the Interdisciplinary Innovation Fund of Nanchang University (Grant No. 2019-9166-27060003), the Open Project of Shandong Provincial Key Laboratory of Optics and Photonic Devices (Grant No. K202102), and the China Scholarship Council (Grant No. 202008420045).

T. L. and Z. H. contributed equally to this work. Z. H. and J. D. received B.S. degree in physics from Nanchang University, and are currently working toward Ph.D. degree at Fudan University.

\end{acknowledgments}


%

\end{document}